\documentclass[runningheads]{llncs}
\usepackage[T1]{fontenc}
\usepackage{graphicx}
\usepackage{booktabs} 
\usepackage{url}
\usepackage{cite}
\usepackage{amsmath}
\usepackage{tabularx}

\begin{document}

\title{Struggle as Flow: Challenge, Design, and Experience in Soulslike Games\thanks{This research was funded by: GR024678 NSERC CREATE 2020 Immersive Technologies, Natural Sciences and Engineering Research Council of Canada; GR026895 SSHRC 2022 Okanagan WaterFutures: Experiential Games for Water Responsibility, Social Sciences and Humanities Research Council of Canada.}}

\titlerunning{Struggle as Flow in Soulslike Games}

\author{Zhehao Sun\inst{1} \and
Yuanyuan Xu\inst{1} \and
Chi Zhen\inst{1} \and
Yin-Shan Lin\inst{2} \and
Miles Thorogood\inst{1} \and
Patricia Lasserre\inst{3} \and
Aleksandra Dulic\inst{1} \and
Megan Smith\inst{1}}
\authorrunning{Z. Sun et al.} 

\institute{Faculty of Creative and Critical Studies, University of British Columbia,\\
Kelowna, BC V1V 1V7, Canada\\and
Department of Computer Science, Irving K. Barber Faculty of Science,\\
University of British Columbia, Okanagan, BC V1V 1V7, Canada}

\maketitle 

\begin{abstract}
While traditional game design prioritizes friction-free accessibility, the Soulslike subgenre has achieved commercial dominance through punishing difficulty and frequent failure. This paper challenges the conventional hedonistic paradigm of gaming to investigate the psychological mechanisms behind the Paradox of Failure. By integrating Csikszentmihalyi's Flow Theory with Juul's ludological framework, we propose the concept of Resilient Flow. We define this as a cognitive state wherein absorption is maintained not despite frustration but through the meaningful framing of it. To validate this model without invasive laboratory constraints, we conducted a qualitative text analysis of 600 helpful user reviews from Elden Ring, Sekiro: Shadows Die Twice, and Dark Souls III via the Steam Community platform. Findings reveal that long-term players linguistically reframe death as pedagogy rather than punishment and utilize vocabulary associated with rhythmic synchronization and meditative focus. We conclude that when difficulty is designed with clarity and fairness, it fosters an Ethics of Attention and transforms digital struggle into a profound experience of mastery and mindfulness.

\end{abstract}

\section{Introduction} Conventional design wisdom suggests that players engage with digital entertainment primarily for relaxation and stress relief. Contradicting this assumption, the Soulslike genre has secured a massive global audience specifically by offering punishing difficulty and frequent failure \cite{guzsvinecz2025}. The market reflects this counterintuitive value proposition: the global Soulslike market is valued at approximately 1.5 billion dollars in 2024 and is projected to reach 3.2 billion dollars by 2033. This financial trajectory implies that difficulty is not a barrier to entry but a primary driver of consumer interest in high-stakes interactive environments \cite{guzsvinecz2023}.
This phenomenon raises a fundamental psychological question regarding why individuals voluntarily persist in environments designed to defeat them \cite{juul2010}. This inquiry connects directly to the concept of Flow, which psychologists define as a state of optimal experience where skill and challenge are perfectly balanced \cite{csikszentmihalyi1990, chen2007}. Traditional models posit that excessive difficulty leads to anxiety and disengagement \cite{deterding2023, abuhamdeh2020}. However, FromSoftware titles such as Elden Ring and Sekiro: Shadows Die Twice maintain player engagement specifically through high-stress encounters that seemingly violate the standard Flow channel by pushing players toward the anxiety boundary to maintain hyper-focus \cite{caracciolo2024}.
This paper argues that high difficulty does not necessarily disrupt the Flow state. Instead, we propose that these games induce a variation of the state we term Resilient Flow. In this model, struggle is not an interruption of enjoyment but the essential material for it \cite{frommel2021}. When setbacks feel fair and rules are consistent, the emotional weight of failure transforms into the satisfaction of mastery \cite{swink2009, hefkaluk2024}. To investigate this dynamic, our research moves beyond theoretical abstraction; we adopt a data mining approach to analyze user reviews from the Steam platform to understand how players linguistically articulate their relationship with digital pain and cognitive absorption \cite{guzsvinecz2023}.

\section{Theoretical Framework} \label{sec:theory}
To understand how high difficulty generates engagement rather than repulsion, it is necessary to synthesize ludological definitions of the Soulslike genre with psychological models of attention and failure. This synthesis allows for an exploration of how the "paradox of failure" functions within digital environments to foster persistence rather than disengagement \cite{hefkaluk2024}. This section outlines the transition from standard Flow Theory to what we define as Resilient Flow by examining the structural role of failure in game design \cite{frommel2021}.
\subsection{The Soulslike Archetype} The Soulslike subgenre has evolved from a niche market into a dominant design philosophy in modern action role-playing games \cite{guzsvinecz2025}. While often reduced simply to high difficulty, the genre is more accurately defined by its specific pedagogical structure \cite{perttula2017}. Caracciolo notes that these games employ narrative opacity and mechanical rigidity to enforce a learning curve \cite{caracciolo2024}. Unlike power fantasy games that empower players through statistical accumulation, Soulslike games empower players through the accumulation of knowledge \cite{mirhadi2024}. The primary mechanic is not the combat itself but the loop of death and rebirth, which transforms the cost of failure into a temporal investment in mastery \cite{juul2010}. When a player dies, they typically lose accumulated resources but retain their knowledge of the level layout and enemy patterns \cite{csikszentmihalyi1990}. This creates a system where the character does not necessarily improve, but the player does, aligning with the "growth principle" where a more complex set of capacities is sought after and developed \cite{admiraal2011}. This distinction is crucial for understanding user engagement; the gameplay loop demands that players actively study the game rather than passively consume it \cite{laurence2023}. Consequently, the high cognitive load required for such active learning blocks out external distractions, facilitating a state of hyper-focus that is essential for maintaining flow in high-stakes environments \cite{abuhamdeh2020}.

\subsection{Flow Theory in High-Stakes Environments}

Csikszentmihalyi originally defined flow as an autotelic state of optimal experience where an individual's skills are fully engaged in overcoming a manageable challenge \cite{csikszentmihalyi1990, admiraal2011}. In the context of game design, this is often visualized as a channel between boredom and anxiety, representing a zone where psychic energy is invested into order in consciousness \cite{chen2007, perttula2017}. Traditional usability principles suggest that designers should act as benevolent guides who smooth out friction and utilize dynamic difficulty adjustment to keep players safely within this optimal channel \cite{fisher2024, kao2024}.

However, Soulslike games challenge this benevolent model by intentionally pushing players toward the anxiety boundary, leveraging high difficulty to force a state of hyper-arousal \cite{deterding2023}. Cowley et al. argue that the flow state in gaming is not static but dynamic, evolving as the player masters the game's rule-bound action sequences \cite{cowley2008}. To maintain high arousal and deep focus, the system must threaten the player with genuine failure; within this framework, the temporal loss of time invested becomes a psychological measure of the difficulty's impact \cite{juul2010}. If the consequences of failure are trivial, attention is allocated elsewhere and intrinsic motivation wanes \cite{abuhamdeh2020, guzsvinecz2023}. By implementing severe penalties for death and demanding high precision, Soulslike games significantly increase the cognitive load required to play, often necessitating that players study narrative clues and mechanics to gain mastery \cite{caracciolo2024}. This high attentional demand leads to "Cognitive Occlusion," a process that effectively crowds out external distractions and intruding negative thoughts by demanding 100\% of the player's resources for immediate survival \cite{ parvizi2024}. Consequently, this high-stakes variation of flow relies on the player's ability to resiliently bounce back from the anxiety edge, transforming potential frustration into a "positive negative experience" characterized by heightened awareness and the satisfaction of overcoming repeated struggle \cite{frommel2021, hefkaluk2024, montola2010}.

\subsection{The Paradox of Failure}
The central psychological anomaly of the genre is what Juul terms the Paradox of Failure \cite{juul2010}. This is the observation that players voluntarily seek out games that make them unhappy in the short term to achieve gratification in the long term, a manifest behavioral pattern where players persist despite setbacks \cite{hefkaluk2024}. Juul argues that failure is central to the aesthetic worth of a game, noting that players generally rate games higher if they fail at least once, as it prompts them to reconsider their strategies and adds perceived depth to the experience \cite{juul2010, frommel2021}. If a game cannot be lost, winning feels inconsequential; therefore, the frustration experienced during a difficult boss fight is not a design flaw but the raw material of the eventual satisfaction.

Montola expands on this by framing these moments as positive negative experiences \cite{montola2010}. The pain of defeat validates the integrity of the game world, fostering an internal locus of control where players attribute failure to their own psychological failings, such as "greed" or "hesitation", rather than system errors. If the game allowed the player to win without effort, it would effectively be lying about the player's competence, which can lead to boredom and a loss of interest once a task is no longer perceived as challenging \cite{deterding2023, frommel2021}. By refusing to compromise its difficulty, the game communicates respect for the player's potential, acting as a "fair pedagogue" that provides consistent and functional feedback \cite{hefkaluk2024}. This aligns with Self-Determination Theory, specifically the need for competence, as overcoming "optimal challenges" is a primary driver of intrinsically motivated behavior and restorative experiences during times of difficulty \cite{mirhadi2024, deterding2023}. When players finally overcome a challenge that successfully stopped them dozens of times, the feedback affirms their mastery in a way that easier games cannot replicate \cite{hefkaluk2024}.

\subsection{Game Feel and the Ethics of Attention}
The bridge between unfair frustration and motivating challenge lies in what Swink defines as Game Feel, which encompasses the tactile, moment-to-moment sensation of controlling a digital avatar \cite{swink2009}. In Soulslike games, movement is often characterized by mechanical rigidity and narrative opacity, where actions are deliberate, weighty, and unchangeable once initiated \cite{caracciolo2024}. This design creates a contract of trust between the player and the system; because the controls are precise and consistent, the player must accept responsibility for their errors rather than attributing failure to external factors \cite{ guzsvinecz2023}.
This relates directly to Deterding’s concept of the Ethics of Attention, suggesting that a game demanding total focus must justify that demand by providing a fair and consistent world \cite{deterding2023}. If a player dies due to a random bug or poor camera angle, the contract is broken, and the flow state collapses into resentment; however, if the player fails because they mistimed a dodge by a fraction of a second, the contract holds and the frustration is directed inward. This pivot toward internal reflection is the specific mechanism that sustains engagement through hundreds of hours of play, transforming the digital struggle from a passive experience of suffering into an active process of learning and eventual mastery \cite{frommel2021, hefkaluk2024}.

\section{Methodology: Qualitative Thematic Analysis} \label{sec:method}
To investigate the phenomenological experience of Resilient Flow, this study adopts a qualitative thematic analysis approach. Instead of relying on laboratory settings which may introduce observation bias, we examined naturally occurring player testimonies from the Steam Community archives. This ecological approach allows for the extraction of psychological states directly from the player base in a naturalistic setting, mirroring methodologies used to assess player opinions in complex RPG environments \cite{guzsvinecz2025}.
The methodological pipeline, visualized in Fig. 1, proceeds through three distinct phases: data construction, inductive coding, and theoretical synthesis.
\subsection{Phase I: Data Sampling and Corpus Construction} Data collection was driven by a purposive sampling strategy. The objective was not to analyze the average player, but the resilient expert who demonstrated a behavioral pattern of persisting through setbacks and high failure counts to achieve mastery \cite{hefkaluk2024}. From an initial corpus of approximately 2,000 entries, we applied a strict filtering funnel to ensure data validity and capture verified player expertise \cite{guzsvinecz2023}.
Specific inclusion criteria were established to filter noise. First, we applied an expertise metric requiring playtime greater than 50 hours \cite{guzsvinecz2023}. Flow states typically occur only when high skill matches high challenge; players with low playtime were excluded to filter out initial frustration bias or lack of development in necessary cognitive capacities \cite{csikszentmihalyi1990}. Second, we selected reviews tagged as "Most Helpful" by the community to ensure the sentiments resonated with the broader player base and filtered out outliers \cite{hefkaluk2024}. Finally, we filtered for informational value, manually removing low-effort submissions or memes in favor of descriptive narratives that articulated the internal cognitive state during high-arousal play \cite{abuhamdeh2020}.

To validate this model without invasive laboratory constraints, we conducted a qualitative text analysis of 600 helpful user reviews from three distinct Soulslike titles: \emph{Elden Ring}, \emph{Sekiro: Shadows Die Twice}, and \emph{Dark Souls III}. These specific titles represent distinct variations of the Soulslike combat loop—rhythmic, reactive, and open-ended—which allow for a comprehensive analysis of narrative complexity and its relationship with player attention \cite{caracciolo2024}.

\subsection{Phase II: The Inductive Coding Procedure}
The analysis followed a hierarchical coding process (see Fig.~\ref{fig:framework}, Col 2), moving from concrete lexical items to abstract psychological dimensions.

First, during open coding, we scanned the corpus for high-frequency lexical descriptors defined by the players themselves. Keywords such as "dance," "rhythm," "greed," and "zen" were isolated as primary descriptors.

Second, during axial coding, these raw keywords were grouped into broader semantic categories. For instance, self-blaming terms like "my fault" were categorized under internal locus of control, while terms regarding "relaxing" were grouped under cognitive occlusion.

Third, during the synthesis phase, these categories were mapped onto existing flow theory frameworks. This abstraction process allowed us to construct the specific dimensions of Resilient Flow discussed in the results.

\subsection{Phase III: Theoretical Output}
\begin{figure}[ht]
    \centering
    \includegraphics[width=1\textwidth]{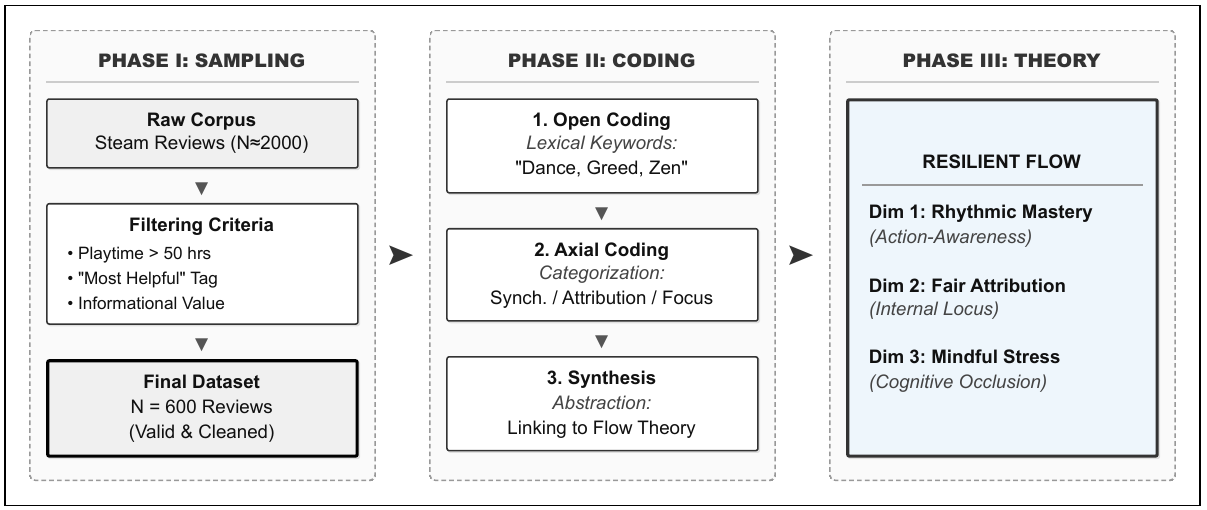}
    \caption{The methodological framework of the study. The diagram illustrates the inductive trajectory from raw data sampling (Phase I), through hierarchical coding procedures (Phase II), to the final theoretical dimensions of Resilient Flow (Phase III).}
    \label{fig:framework}
\end{figure}

The final stage of the methodology integrated the coded data into a cohesive theoretical model (see Fig.~\ref{fig:framework}, Col 3). This process distilled the player experience into three structural dimensions: rhythmic mastery, algorithmic fairness, and mindful stress. This structured approach ensures that the findings presented in the subsequent analysis section are empirically grounded in the collective consciousness of the player community rather than anecdotal evidence.

\section{Analysis: The Collective Voice of the Player Structure}
\label{sec:analysis}

The qualitative analysis of the Steam Community corpus (N=600 reviews across three titles) reveals that player experiences are not monolithic but evolve through distinct phases. By aggregating individual testimonials, we identified a shared lexicon used to describe the transition from frustration to mastery. The analysis is structured into three dimensions: the sensory metaphor of combat, the psychological attribution of failure, and the therapeutic impact of the environment.

\subsection{Dimension 1: The Metaphorical Shift (Violence vs. Performance)}
A semantic frequency analysis reveals a distinct linguistic shift within the players' retrospective narratives. When describing their initial encounters with the game mechanics, users employ terminology of violence (e.g., ``beating,'' ``killing,'' ``surviving''). In contrast, as they articulate their current state of mastery (50+ hours), the lexicon shifts to the terminology of performance art and music.

Table~\ref{tab:linguistic_shift} categorizes these descriptors, highlighting how the perception of combat evolves from a struggle for survival into a rhythmic performance.

\begin{table}[hbt!]
\caption{Linguistic Transformation of Combat Descriptions in Player Reviews}\label{tab:linguistic_shift}
\centering
\renewcommand{\arraystretch}{1.3}
\setlength{\tabcolsep}{6pt}
\begin{tabular}{|l|p{0.35\linewidth}|p{0.45\linewidth}|}
\hline
\textbf{Game Title} & \textbf{Phase 1: Initial Encounter} & \textbf{Phase 2: Adapted State} \\
\hline
\emph{Sekiro: Shadows Die Twice} & ``Brutal,'' ``Impossible,'' ``Suffering,'' ``Getting minimal progress.'' & ``Rhythm game,'' ``A violent waltz,'' ``Dance with them,'' ``Clicking,'' ``Music of swords.'' \\
\hline
\emph{Dark Souls III} & ``Punishing,'' ``Unfair,'' ``Clunky,'' ``Dying over and over.'' & ``Formal duel,'' ``Exchange,'' ``Turn-based combat,'' ``Respect the mechanics.'' \\
\hline
\emph{Elden Ring} & ``Overwhelmed,'' ``Lost,'' ``Panic rolling,'' ``Getting destroyed.'' & ``Choreography,'' ``Spectacle,'' ``Cinematic fight,'' ``Beautiful dance.'' \\
\hline
\end{tabular}
\end{table}

The most vivid evidence comes from \emph{Sekiro: Shadows Die Twice} players who cease to see enemies as antagonists and begin to view them as dance partners. One review explicitly states:
\begin{quote}
``The gameplay is on a whole different level... once you master the parry, you don’t fight enemies anymore... you dance with them. The combat becomes a violent waltz.'' (Steam User, 126 hours)
\end{quote}
Another user reinforces this auditory-tactile link:
\begin{quote}
``There is no better noise in my life than cling and clang... Even the deflecting sound, it gives me such an orgasm.'' (Steam User, 313 hours)
\end{quote}

\subsection{Dimension 2: The Taxonomy of Failure (Why We Die)}
Contrary to the assumption that high difficulty leads to resentment, the reviews demonstrate a strong Internal Locus of Control. Players do not blame the system; they blame a specific set of psychological failings. We categorized over 200 descriptions of ``death'' into a taxonomy of error, as shown in Table~\ref{tab:failure_taxonomy}.

\begin{table}[hbt!]
\caption{Comparative Psychological Profiles of the Three Titles}\label{tab:psych_profiles}
\centering
\renewcommand{\arraystretch}{1.3} 
\setlength{\tabcolsep}{6pt}
\begin{tabular}{|l|p{0.28\linewidth}|p{0.28\linewidth}|p{0.28\linewidth}|}
\hline
\textbf{Feature} & \textbf{\emph{Sekiro: Shadows Die Twice}} & \textbf{\emph{Dark Souls III}} & \textbf{\emph{Elden Ring}} \\
\hline
\textbf{Primary Stressor} & Speed \& Precision & Spacing \& Stamina & Scale \& Unknown \\
\hline
\textbf{Flow State} & \textbf{Hyper-Focus}: "Mushin" (No-Mind), instinctual reaction. & \textbf{Rhythmic Endurance}: Managing resources in a grim world. & \textbf{Wonder}: Flow interrupted by moments of awe and discovery. \\
\hline
\textbf{Player Sentiment} & "I feel like a master swordsman." & "I feel like a survivor in a dying world." & "I feel small in a majestic universe." \\
\hline
\end{tabular}
\end{table}

The concept of ``Greed'' is particularly prominent in \emph{Dark Souls III}, representing a moral failure of the player rather than a mechanical failure of the character. As one user reflects:
\begin{quote}
``I got too greedy trying to get another swing in and it cost me... The difficulty is sometimes unfair but 99\% of my deaths are my fault.''
\end{quote}
This acceptance of blame turns the game into a ``fair pedagogue,'' where punishment is consistent and predictable.

\subsection{Dimension 3: The Therapeutic Paradox}
Perhaps the most surprising finding from the dataset is the high frequency of players describing these stressful games as ``relaxing'' or ``depressurizing.''

In the \emph{Elden Ring} and \emph{Dark Souls III} reviews, players frequently cited the game as a remedy for real-life anxiety. The high cognitive load required to play forces a state of cognitive occlusion, where the brain is too busy surviving to process external stressors.

One user explicitly connects the difficulty to mental health relief:
\begin{quote}
``I was in a very dark place in my life, full of depression... This game taught me patience and perseverance. It put me in a state where the only thing that mattered was the present moment.'' (Steam User, \emph{Dark Souls III})
\end{quote}

Another player describes \emph{Sekiro: Shadows Die Twice} as a form of meditation:
\begin{quote}
``I bought Sekiro: Shadows Die Twice to relax after work... I’ve achieved a kind of spiritual enlightenment that only comes from being repeatedly stabbed by a seven-foot monkey... It clears your head like nothing else.''
\end{quote}

\subsection{Cross-Game Psychological Profiles}
Finally, we synthesized the data to construct a comparative psychological profile for each title within the Soulslike genre. While they share a mechanical foundation, the player's emotional journey differs significantly, as summarized in Table~\ref{tab:psych_profiles}.

\begin{table}[hbt!]
\caption{Comparative Psychological Profiles of the Three Titles}\label{tab:psych_profiles}
\centering
\renewcommand{\arraystretch}{1.3} 
\setlength{\tabcolsep}{3pt}       
\begin{tabular}{|l|p{0.28\linewidth}|p{0.3\linewidth}|p{0.3\linewidth}|}
\hline
\textbf{Feature} & \textbf{\emph{Sekiro: Shadows Die Twice}} & \textbf{\emph{Dark Souls III}} & \textbf{\emph{Elden Ring}} \\
\hline
\textbf{Primary Stressor} & Speed \& Precision & Spacing \& Stamina & Scale \& Unknown \\
\hline
\textbf{Flow State} & \textbf{Hyper-Focus}: "Mushin" (No-Mind), instinctual reaction. & \textbf{Rhythmic Endurance}: Managing resources in a grim world. & \textbf{Wonder}: Flow interrupted by moments of awe and discovery. \\
\hline
\textbf{Player Sentiment} & "I feel like a master swordsman." & "I feel like a survivor in a dying world." & "I feel small in a majestic universe." \\
\hline
\end{tabular}
\end{table}

\section{Discussion: The Pedagogy of Suffering}
\label{sec:discussion}

The qualitative evidence gathered from the Steam Community Archive suggests that the appeal of Soulslike games extends beyond simple masochism or the desire for bragging rights. Instead, these games construct a rigorous psychological framework that we term Resilient Flow. This state is distinct from traditional Flow in that it incorporates failure as a necessary rhythmic beat rather than a disruption.

\subsection{Redefining the Flow Channel}
Classic Flow Theory suggests that anxiety occurs when challenge exceeds skill, typically leading to disengagement. However, our findings indicate that in Soulslike games, the Flow channel is not static. By framing combat as a ``dance'' or ``rhythm'' (particularly in Sekiro: Shadows Die Twice), the game design invites players to linger on the very edge of anxiety. The high-frequency demands of the gameplay force the player into a state of hyper-arousal. Unlike the relaxing state of low-stakes gaming, this hyper-arousal creates a sharper form of focus. The player does not simply react; they synchronize. The transition from describing enemies as obstacles to describing them as dance partners signifies a fundamental shift in perception: the threat is no longer an external stressor but a cue for performance.

\subsection{Fairness as the Gatekeeper of Agency}
The analysis of the term ``Greed'' in Dark Souls III reviews reveals the critical role of fairness in maintaining player agency. Juul's Paradox of Failure holds true only when the failure is perceived as valid. If a game is buggy or inconsistent, failure leads to resentment. However, specific design heuristics in FromSoftware titles, such as telegraphed attack animations and consistent hitboxes, ensure that death is perceived as a logical consequence of user error. By attributing failure to personal vices like ``greed'' or ``panic,'' players maintain an Internal Locus of Control. This psychological mechanism protects the player from helplessness. As long as the player believes the error was internal, they believe it can be corrected, thus sustaining the motivation to re-engage with the loop of suffering.

\subsection{Cognitive Occlusion and Therapeutic Stress}
Perhaps the most significant contribution of this study is the validation of high-difficulty gaming as a mindfulness practice, as evidenced by the Elden Ring dataset. The concept of Cognitive Occlusion suggests that the human brain has a limited processing bandwidth. By demanding 100\% of the player's cognitive resources for immediate survival, Soulslike games effectively crowd out intrusive negative thoughts related to real-world anxieties. The ``Zen'' state described by players is not the peace of inaction, but the peace of singular action. In a world of complex, ambiguous real-life stressors, the binary clarity of the Soulslike contract, survive or die, provides a paradoxical psychological relief.

\section{Conclusion}
\label{sec:conclusion}

This paper challenged the hedonistic assumption that video games must provide friction-free empowerment to be enjoyable. Through a qualitative analysis of 600 player reviews across Elden Ring, Sekiro: Shadows Die Twice, and Dark Souls III, we identified the psychological architecture of Resilient Flow. We conclude that high difficulty transforms from a barrier into a gateway for optimal experience when three conditions are met: rhythmic synchronization, transparent fairness, and total cognitive absorption.

\subsection{Summary of Contributions}
Our analysis demonstrates that successful difficult games function as pedagogical systems. They teach players to reinterpret failure not as a stop state, but as information.
\begin{itemize}
    \item In Sekiro: Shadows Die Twice, we observed how the metaphor of the ``dance'' allows players to process high-speed violence as aesthetic rhythm.
    \item In Dark Souls III, we found that the concept of ``fairness'' allows players to internalize failure as ``greed,'' thereby maintaining their sense of agency.
    \item In Elden Ring, we uncovered how the scale of difficulty induces a ``Zen'' state, providing therapeutic relief through cognitive occlusion.
\end{itemize}
Collectively, these findings suggest that the Soulslike genre succeeds because it respects the player's capacity for growth, offering a sense of earned mastery that automated power fantasies cannot replicate.

\subsection{Limitations and Future Work}
This study relied on text-based self-reports from a specific subset of players, those who were motivated enough to write positive or constructive reviews on Steam. This introduces a survivorship bias; we did not analyze the psychological state of players who quit the game in frustration and never returned. Furthermore, textual analysis cannot measure physiological markers of flow.

Future research should look to integrate biometric data (such as heart rate variability and skin conductance) to correlate these reported psychological states with physiological arousal during boss encounters. Additionally, comparative studies between the Soulslike genre and other high-difficulty genres, such as roguelikes or competitive shooters, could further illuminate the universality of Resilient Flow. Ultimately, understanding why we choose to struggle in digital worlds may provide deeper insights into human resilience in the real world.

\section*{Acknowledgements}
This research was funded by: GR024678 NSERC CREATE 2020 Immersive Technologies, Natural Sciences and Engineering Research Council of Canada; GR026895 SSHRC 2022 Okanagan WaterFutures: Experiential Games for Water Responsibility, Social Sciences and Humanities Research Council of Canada.


\end{document}